\documentstyle[aps,prb]{revtex}
\begin{document}
\draft
%\twocolumn[\hsize\textwidth\columnwidth\hsize\csname @twocolumnfalse\endcsname
\title{Notes on Decoherence at Absolute Zero}
\author{Pritiraj Mohanty \\
        {\it Condensed Matter Physics 114-36, California Institute of Technology, 
        Pasadena, CA 91125}\\
	Email: mohanty@caltech.edu\\
	(To be published in Physica B, LT22 proceedings)
%\date{\today}
}
\maketitle

\begin{abstract}
The problem of electron decoherence at low temperature is analyzed from the 
perspective of recent experiments on decoherence rate measurement
and on related localization phenomena in low-dimensional systems.
Importance of decoherence at zero temperature, perhaps induced 
by quantum fluctuations, is put in a broader context. 

\end{abstract}

% write here 3 or 4 keywords separated by semicolons

\section{Introduction}

Decoherence is the process which---through the interaction of the system with external
degrees of freedom referred to as an environment---sustains a loss of quantum coherence 
in a system. It defines the transition from quantum behavior of a closed system, 
which thus possesses unitarity or time reversibility and displays interference due to the 
superposition of its wave function, to the classical behavior of the same as an open 
system; the loss of unitarity or time-reversal symmetry  
leads to a loss of interference\cite{zurek}. This openness comes from the 
coupling of the  quantum system to an environment or a 
bath\cite{caldeira-leggett,Joos-Zeh}. A closed system,
on the other hand, does not undergo decoherence. The quantum system in question could 
be an electron  whereas the environment could be thermal phonons or photons, 
and even other electrons whose properties are not measured. The coarse-graining of the 
irrelevant degrees of freedom defining the environment, which are not of interest to 
the measurement, generates both dissipation and decoherence: the latter formally 
related to the decay of the off-diagonal terms of the reduced density matrix operator 
denoting the quantum system.

The interpretational problem with decoherence, and in fact the notion of 
decoherence itself,  vanishes when one treats the system-environment
combination as one indivisible quantum object. The combination is closed and evolves
unitarily according to the laws of quantum mechanics transforming pure states into
pure states, hence there is no decoherence. 
The problem only arises in the splitting of the whole as ``a system of interest'' to the 
observer or the experiment, and the remaining degrees of freedom as ``the environment''. 
This split is necessary and must be acknowledged from the observer's or experiment's 
perspective. Interestingly, a pure
state of the closed combination is compatible with each part being in mixed
states. Decoherence is obtained by considering the density matrix operator for 
the combination and partially tracing out the irrelevant degrees of freedom, namely those
of the environment. The reduced density matrix operator then represents the ``effective''
system alone as a statistical mixture, which is of interest to a measurement in an 
experiment. An initially isolated system inevitably loses quantum coherence due to 
its coupling to a complex or a ``large'' environment with very many degrees of freedom.
When both the system and the environment are treated quantum mechanically, the quantum
entanglement becomes an important concern for the loss of coherence.

The loss of coherence of an electron inside a disordered conductor occurs due 
to the interaction with environments: its coupling to localized spins---pseudo 
or magnetic, electron-phonon interactions and electron-electron interactions, 
the latter being  dominant at low temperature. Conventional theories\cite{aa85}
decree that the suppresion of coherence, characterized by a decoherence rate 
$1/\tau_\phi$ vanish with decreasing temperature, ultimately giving 
a Fermi-liquid ground state.  However, in experiments a finite decoherence rate
is observed at low temperatures\cite{prl97}, which perhaps persists down to $T=0$. 
Considering the consequences of such an observation, to be discussed in sections 3 and 4,
it is imperative to put the experimental observation on firm ground. Towards that end, 
our experimental observation of $\tau_\phi$ saturation has undergone extensive 
experimental checks
detailed in section 2. Corroborative problems in mesoscopics denoting severe
discrepancies between experiments and the conventional theories are outlined
in section 3; a connection between these discrepancies and $\tau_\phi$ saturation 
is made. In the final section, zero temperature decoherence and the role
of quantum fluctuations of the environment is put in a broader perspective.

It is argued that zero temperature decoherence observed in low-dimensional
electronic systems is important in understanding various low temperature properties
of metals, acceptance of which as an intrinsic effect appears imminent.

%_____________________________________________
\section{Electron and its environments: \hfill 
Measurement of electron decoherence rate}
%______________________________________________
Inside a disordered conductor, an electron undergoes various kinds of interference. 
The interference of two paths in a doubly-connected regime gives an Aharonov-Bohm 
correction to the electron conductance, which can be modulated periodically 
as a function of the applied field.
Similarly, interference correction arising from paths inside a
conductor in a singly-connected regime gives reproducible conductance fluctuations. 
If the interfering paths are a time-reversed pair, then the correction to the conductance
gives weak localization which can be suppressed by the application of a magnetic field. 
Persistent current is also observed due to interference in isolated metal rings.  

Interference due to phase coherence in the electron wave function can be studied 
using any of these effects if the exact dependence of the measured quantity can be
explicitly expressed in terms of a decoherence rate $1/\tau_\phi$. 
Weak localization correction,
though the least exotic of the effects mentioned above, gives a single parameter
estimate of decoherence rate $1/\tau_\phi$ without any further assumption regarding
the effect. Physically, it is meaningful then to imagine the breaking of
time-reversal symmetry and the emergence of non-unitarity as the suppression of 
interference between the time-reversed paths by an applied magnetic field.

%______________________________________________________
\begin{figure}
 \vbox to 9cm {\vss\hbox to 7cm
 {\hss\
   {\includegraphics{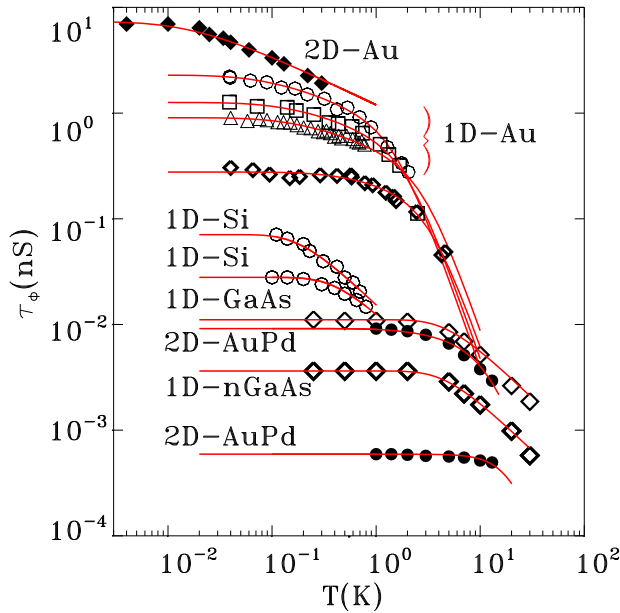}
   }
  \hss}
 }
\caption{Measured decoherence time in various mesoscopic systems.}
\end{figure}
%_____________________________________________________________________________

Fig.~1 displays a small representative of a vast body of data available in the 
literature. What is observed in the experiments is the following: 
(a) At high temperatures the decoherence rate $1/\tau_\phi$ is temperature 
dependent due to various mechanisms such as electron-phonon and electron-electron 
interactions, but at low temperatures the rate
inevitably saturates, suggesting the onset of a temperature independent mechanism. 
(b) The limiting rate $1/\tau_0$ and the temperature at which it dominates vary
over a wide range depending on the system, though a one-to-one correlation with
the sample parameters such as the diffusion constant $D$, or resistance per unit 
length $R/L$ can be made very accurately\cite{prl97,prb97,zaikin}.

A compilation of some saturation data in various systems 
is contained in ref.\cite{prl97,zaikin}. In view of these experiments
it seems plausible that the observed saturation could be a real effect. 
Such a hypothesis must be thoroghly investigated, since the saturation of 
decoherence rate suggesting an intrinsic decoherence is known to have serious 
consequences. To that end, we have performed various control experiments which 
suggest that this limiting mechanism is not
due to any artifacts and is intrinsic. Extensive checks for the role
of various artifacts include the following:

\subsection{\it Heating of the system:} 

Loss of thermal contact of the electron in the sample with the cryostat 
would imply that the temperature of the sample is 
locked at the apparent saturation temperature $T_0$. In the experiment, 
the electron temperature was determined by measuring the electron-electron 
interaction(EEI) correction to the conductivity\cite{aa85} at a magnetic 
field strong enough to quench weak localization. Electron temperature
was found to be in equilibrium with the cryostat to within 
a temperature\cite{prl97} of an order of magnitude less than $T_0$. 

\subsection{\it Magnetic impurities:}

Magnetic impurities such as iron(Fe) in a host metal of gold(Au) were shown 
not to cause saturation\cite{prl97}, contrary to an earlier notion and 
consistent with other experiments\cite{bergmann}. A detailed 
study\cite{mw-kondo} revealed interesting properties of Kondo systems in 
quasi-1D systems, different from the anticipated behavior for 
the bulk Kondo systems.  

\subsection{\it External high-frequency noise:}

Initial checks\cite{prl97}, confirmed by subsequent controlled experiments,
showed that externally generated high frequency(HF) noise\cite{aga} did not 
cause dephasing before heating the sample\cite{webb-hf} to a substantially 
higher temperature. A similar control experiment 
on the saturation of $1/\tau_\phi$ in quantum dots reached the same 
conclusion\cite{marcus-hf}.

\subsection{\it Two-level systems:}

Recently an argument\cite{imry2} was made that nonmagnetic
impurities, which in principle give rise to a dynamic
or time-dependent disorder, could be responsible for the observed saturation;
such defects, usually modeled as two-level systems (TLS), result in the usual
low-frequency $1/f$ noise in conductors. For the following reasons
TLS can be ruled out as the effective environment in our experiments:
(a)A typical level
of noise power of $10^{-15} W$ at $\sim$ 1 GHz ($\omega \sim \tau_0^{-1}$), 
required for dephasing\cite{aga,webb-hf}, would suggest a power level of 1 $\mu W$
or higher at low frequencies (1 mHz-10 Hz). At such high power levels one would 
anticipate the observation of low-frequency switching or hysteresis. Neither
phenomenon was observed in our experiment on timescales of months.   
(b)Another reason for the TLS to be ineffective in our gold samples is
the signature of mesoscopic dimensions in the temperature dependence, contrary
to an expected bulk dependence as in any ``Kondo-like'' theory.  
(c) In the model\cite{imry2}, $\tau_\phi \propto T^{-1}$ in the 
temperature dependent regime,
whereas in the experiment $1/\sqrt{T}$ dependence was observed\cite{prl97} 
for most of the metallic samples.

Another construct based
on the presence of dynamical nonmagnetic impurities or TLS\cite{ralph} suggests 
that the coupling between the TLS and the electron in a metal could give  finite
scattering even at T=0 in the non-Fermi-liquid regime, i.e. below the corresponding
Kondo temperature $T_K$. In this clever construct
it is expected, above and beyond the anticipated behavior of TLS discussed
earlier, the observed saturation rate will be non-unique and history 
dependent.  
But no dependence on history or on annealing was observed over a period of 
months in our experiments. For these reasons 
two-level systems are not thought to be relevant to our observed saturation.

\subsection{\it Openness to external phonons in the leads:}

This nonequilibrium effect arises because of the contact leads to the sample,
necessary for measurement. It has been suggested\cite{open-phonon}, based
on earlier arguments\cite{landauer}, that due to electron-phonon coupling 
phonons in the leads exist as an inevitable extrinsic environment. 
The associated phonon emission process
gives an effective lifetime to the electron. It is argued that low temperature 
saturation is determined by the contact geometry and configurations, and the  
dependence at high temperature  is determined by material properties. 
First, in our experiments, in anticipation of such a possibility, 
the 2D contact pads were fabricated
at least a length of 3-5 $L_\phi$ away from the four-probe part of the sample. 
Leads to the 2D pads of this length had the same geometry 
as the sample itself. The effect of 2D pads in the weak localization
traces was not detected, and the traces were very different
from the 2D weak localization functional form.

For the high temperature part, 
a large body of data compiled in ref.\cite{prl97,zaikin} shows the lack of material 
dependence of $\tau_\phi$. The dependence\cite{open-phonon} 
is due to very different diffusion constants and other sample parameters 
in the systems compared. Finally,
description of the saturation value in terms of only intrinsic parameters of the 
sample\cite{prl97,prb97,zaikin} argues against the effectiveness of the proposed
mechanism in our experiments.

\subsection{\it Other artifactual environments:}

There are suggestions that gravity\cite{ben-jacob} is
an inevitable environment, making every system essentially open. 
It has also been argued that the nuclear magnetic moment of gold--representing
nuclear degrees of freedom--may  provide an effective environment for
temperature independent decoherence. The last two
suggested mechanisms have been ruled out by incorporating experiments on different
materials, and by the observation of the obvious parametric size dependence
in the same material(Au)\cite{prl97}.

After considering most of the extraneous effects it was concluded that the observed
saturation of $\tau_\phi$ in our experiments is an intrinsic effect.

\section{Manifestation of zero temperature decoherence in mesoscopic physics}

If the premise is assumed, for the sake of arguments in this section, 
that the temperature
independent dephasing of electrons is intrinsic, then the saturation of $\tau_\phi$
must manifest itself  ubiquitously in low-dimensional electron systems by behavior
including but not limited to low temperature saturation of the appropriate
physical quantity.

%_______________________________
\subsection{\it Saturation of $\tau_\phi$ in all dimensions:} 

The data in Fig.1 show saturation of $\tau_\phi$ in quasi-1D and 2D disordered
conductors. Recent experiments report the observation of saturation in $\tau_\phi$
in open ballistic quantum dots, representing 0D systems, below a temperature of 100 mK 
in one set of experiments\cite{marcus-hf}, and below 1K in another set of 
experiments\cite{pivin}. $\tau_\phi$  in 3D amorphous  
Ca-Al-X (X=Au,Ag) alloys also saturates below 4K\cite{olsen}. 
%_______________________________

\subsection{\it Other manifestations in quasi-1D:   
persistent current and e-e interaction:}

It is understood that the saturation of $\tau_\phi$ would imply
a similar saturation in the electron-electron interaction (EEI) correction\cite{schwab-2} 
to the conductivity, measured at a finite field with the weak localization
contribution quenched. The saturation temperature for EEI correction should 
be lower, and comparable to $\hbar/\tau_0$. Experiments\cite{webb-ee} do show
such a saturation and a strong correlation between the EEI saturation temperature 
and $\tau_0$.

Saturation of $\tau_\phi$ also offers solution to the problem of persistent current 
in normal metals\cite{annalen}, namely that the 
observed current is too large and diamagnetic. In experiments,
the range of temperature in which a persistent current is measured is indeed
the same where $\tau_\phi$ is saturated. 
Intrinsic high-frequency fluctuations---responsible for $\tau_\phi$ 
saturation---will 
imply the presence of a non-decaying diffusion current, corresponding to 
the persistent current with a size comparable to $e/\tau_D \equiv eD/L^2$.

%________________________________
\subsection{\it Transition from weak-to-strong localization in quasi-1D conductors:}

A finite decoherence rate at zero temperature is expected to stop the 
Thouless transition\cite{thouless} from weakly to strongly 
localized states. This disorder-driven transition to localized states in 
quasi-1D, with the characteristic length scale of $\xi$ has two possible courses 
depending on the competing length scale of diffusion characterized by 
$L_\phi$ at $T=0$: (i) Complete suppression (no transition at 
all, $L_0 \ll \xi$); (ii) Inihibition (activation with 
decreasing temperature--denoting a transition to a strongly localized state--
inevitably saturates: $L_0 \sim \xi$ in the experimental range). Both aspects 
have been well documented in experiments on $\delta$-doped GaAs wires\cite{misha}
and GaAs-Si wires\cite{sanquer}.

%________________________________
\subsection{\it Lack of one-parameter scaling:} 

One-parameter scaling theory of localization\cite{abrahams}, the foundation
for the theory of low-dimensional conductors, requires phase coherence length
to diverge as a negative power of $T$: $L_\phi \propto T^{-p/2}$. A finite 
temperature-independent decoherence length $L_0$ immediately suggests 
breakdown of the one-parameter scaling theory. Experiments on Si-MOS systems 
have convincingly shown\cite{pudalov-2} the lack of one-parameter scaling.

%________________________________
\subsection{\it Metallic behavior in 2D systems:}

In contrast to the conventional theory of metals\cite{abrahams} which purports
that 2D systems at $T=0$  become insulators with zero conductivity, recent 
experiments\cite{kravchenko} find metallic behavior at low temperatures.
Furthermore, at low temperatures the conductivity of the metallic state
is observed to saturate with a finite value\cite{pudalov-1}. 
However, a nonvanishing decoherence of the electron  
would suggest finite diffusion of the electron, and hence no Fermi liquid 
ground state or insulating state with zero conductivity at $T=0$. With 
decreasing temperature, localization driven by disorder is suppressed, 
sometimes even before the onset by zero-temperature dephasing, depending
on the competition.

Formation of insulating states is inhibited by diffusion induced
by the zero temperature decoherence, irrespective of the initial states.
The quantum-hall-to-insulator transition is in some sense similar to the 
transition in quasi-1D or 2D conducting systems. A quantum-hall system 
beyond a critical field $B_c$ becomes insulating with a diverging 
$\rho_{xx}$ as T is reduced\cite{paalanen}. However, formation of this 
insulating state is expected to be inhibited with a low $T$ saturation of 
the increasing $\rho_{xx}$. Such a saturation has been observed\cite{shahar},
and, on the basis of a recent theory\cite{pryadko}, it is related to a finite
dephasing length at low T; This may perhaps be the size of the puddle  in the 
quantum-hall liquid. Likewise in superconductor-to-insulator transition in 2D a-MoGe
films a similar leveling of the resistance was observed\cite{yazdani} 
with the conclusion that the saturation is due to the coupling to a 
low temperature dissipative environment\cite{mason}.

\section{Counterpoint to conventional theories}

The conventional theory of metals, specifically in low dimensions, is based
on the scaling laws of localization\cite{abrahams} augmented by the perturbative 
treatment of interaction\cite{aa85}. The very nature of these 
theories requires that the phase coherence length diverge with decreasing 
temperature according to a power law, $L_\phi \propto T^{-p/2}$, for some 
positive $p$. The early phenomenological motivation of such a diverging form 
at low T was formalized
in a perturbative calculation of dephasing length\cite{aak,aa85}. The structure of 
the Fermi liquid picture, that the electron interaction can be treated
as low-lying excitations of a non-interacting system while maintaining the 
Fermi liquid ground state at $T=0$, is fully retained even in the presence of disorder
at low dimensions.  

Our experimental observation of $\tau_\phi$, or equivalently $L_\phi$, 
saturation argues against the premise of the conventional theory, and it contradicts
the supporting theory\cite{aak} of electron dephasing in low dimensions. In the last two
sections, a phenomenological case is made against the premise of the conventional
theory. In the following
we briefly discuss the lack of validity of these theories at low temperatures.

Let us just consider the electron-phonon interaction for the sake of
argument. In a conductivity experiment only the scattering rate of the electron
is measured, which includes electron-phonon scattering. Traditionally, the 
relevant phonon states available for an electron to scatter off depends on
temperature $T$ via thermal population. As $T \rightarrow 0$, this population
shrinks to zero, making the scattering rate of the electron  vanish.
By this argument most scattering mechanisms yield
vanishing scattering rate at $T=0$, where the states to be scattered off are thermally
populated. Non-thermal scattering processes obviously do not have to vanish
at $T=0$. The phase shift in the electron wave function $\delta\phi$ 
arising out of electron scattering, say off the phonons in a phonon bath, is random 
($\langle \delta\phi \rangle = 0$) and on averaging it produces a dephasing effect as
a suppression of the interference term by a factor 
$e^{-{t/\tau_\phi}} \equiv \langle e^{i\delta\phi} \rangle$. This indicates that (a) 
phase shifts arise only in presence of a thermal population, and (b) the bath of
phonons itself does not undergo any change which might have an effect or 
back reaction on the electron; in other words, there is no entanglement between 
the electron and the bath. The last two statements are often phrased differently: 
the electron acquires phase shifts due to its coupling to equilibrium fluctuations
of the bath, and the vanishing population through which T enters in the equation 
has to satisfy the law of detailed balance.

This is a point of view, and a limited one at best, for the following reason.
If one starts with a ground state of the electron and the ground state of the 
environment and the coupling is turned on, then the product state evolves in such 
a way, even at zero temperature, that after a certain time the electron is no more 
in its ground state entirely; there is a fractional probability of finding the
electron in its ground state. In other words, the electron can be described only by
a mixed state of both the environmental variables and the electron variables.
The electron can be measured only after the integration of the irrelevant 
environmental variables, the very process that introduces decoherence.   

What is measured in above-mentioned experiments is not a property of the 
combined system of the electron and environment. In the measurement process the 
environmental degrees of freedom are averaged out, the effect of averaging is still
retained in the measured quantity. Thus the electron cannot be considered to be
a closed system, and the notion of a unique ground state in such a case is 
meaningless.

An electron must exhibit zero temperature decoherence if it is
coupled to a phonon bath; the problem is isomorphic to the Caldeira-Leggett model
which does indeed show zero temperature decoherence. The same is true for 
an electron coupled to a fluctuating electromagnetic field, representing 
electron-electron interaction, in spite of complications due to the Pauli exclusion
principle. To summarize the case against conventional theories, (a) experimental
evidence is overwhelmingly against, (b) a quantum mechanical treatment
of the problem does give agreeable results, (c) certain other outstanding problems
can be understood with the notion of zero temperature decoherence, and finally 
(d) the basic theory of decoherence in an exactly solvable model of Caldeira-Leggett 
is contrary to the conclusions of these theories\cite{weiss}.

\section{Endnotes: Quantum fluctuations and decoherence}

To explain the results of the experiments\cite{prl97}, it was 
suggested\cite{prb97} that high-frequency fluctuations of quantum 
origin could indeed cause the saturation. Following 
the well-established concept\cite{aak}, of dephasing of an 
electron by ``classical'' electromagnetic field fluctuations, it is reasonable 
to consider decoherence due to the coupling of the electron to
quantum fluctuations of the field. Such an extension is not new, and is
well known in quantum brownian motion\cite{caldeira-leggett}. A particle coupled
linearly to a bath of oscillators, all in their individual ground states, with a
linear coupling, shifts the equlibrium position of individual oscillators 
without exciting them. The resulting back reaction on the particle causes
both dissipation and decoherence even at absolute zero, the latter 
quantified by the decay of off-diagonal elements of the reduced density matrix 
in the long time limit\cite{caldeira-leggett,unruh,sinha,zaikin-2}. 
A similar construct has been made earlier\cite{loss} in the mesoscopic context. 
The cut-off dependent result is universal in the mesoscopic models as well as 
in quantum brownian motion models. 

The initial back-of-the-envelope
calculation\cite{prb97} surprisingly described the saturation rate
observed in many experiments. The rigorous and commendable calculations\cite{zaikin} 
which verified the notion have been severely criticized\cite{aag,vavilov}. Though the
latter calculations are self consistent\cite{aag}, the theories fail
at the starting point. A pedestrian argument against the use of the ``law
of detailed balance''\cite{aag,imry2} is that it describes only thermal 
transitions. To understand zero temperature effects one must add a non-thermal part, 
put in by hand, as is normally done for spontaneous emission in the Einstein rate 
equation for a laser. 

As mentioned in the Introduction, it is the entanglement of 
the environment with the electron that contributes to the decoherence even though 
energy exchange is not allowed between the individual  non-interacting  
parts, i.e. the electron and the electromagnetic field modes. The combination
is a closed system and does evolve unitarily without decoherence, but the individual
parts can remain in mixed states at the same time. In terms of photons, 
one can imagine exchange of  virtual pairs of photons with the field by the electron
along two different interfering paths. Such an interpretation is often misunderstood
as dressing of an electron or an atom by vacuum fluctuations, and is often a source
of confusing debate. 

There have been a few parallel developments surrounding the
question--whether or not quantum fluctuations can cause decoherence. The role of
vacuum fluctuations in decohering atomic coherence has been discussed 
recently\cite{santos}. The decoherence of an electron due to its coupling to 
vacuum fluctuations has also been previously considered\cite{braginsky} with 
an affirmative conclusion. There was another interesting development on the problem 
of a quantum limit of information processing pertaining to  computation. Starting
from a well known result from black hole entropy theory, a proposal
was made suggesting  quantum-limited information loss\cite{beckenstein}, 
quantified by entropy. This was severely 
criticized\cite{deutsch} again with the argument that zero-point energy cannot be
dissipated as ``heat''. Though the debate was unresolved\cite{bremermann}, since then
it is known in the refined description of decoherence\cite{unruh} that a part of the entropy 
can reside in the correlation. The sum total of entropy of a system, 
bath and that contained in the correlation is equal to the entropy of the combination. 
This is a different way of saying that a pure state of the combination is consistent 
with partial mixed states. All these above-mentioned debates were not settled due 
to the lack of any experiments. Fortunately, our problem starts from experimental 
results.

In conclusion, our experiments along with almost all existing experiments on the direct
or indirect measurement of decoherence rate
are more than suggestive of a non-thermal mechanism, which is in all probability
intrinsic. Existence of field fluctuations at frequencies higher than the
temperature, irrespective of their origin, can  explain various
discrepancies in mesoscopic physics.
In this paper we briefly discussed how persistent current and electron-electron 
interaction correction may be affected by the
saturation of decoherence rate. Following similar arguments, the experimentally observed 
formation of metallic states in 2D, lack of universal one-parameter scaling, 
suppression or saturation of strong localization, and suppression of 
quantum-hall-insulator transition can be understood. 
In mesoscopic physics alone, the fundamental 
role of low temperature behavior of electron decoherence cannot  be 
overemphasized. 

I acknowledge experimental collaboration with R.A. Webb and E.M.Q. Jariwala,
and the formal support of M.L. Roukes.

\end{document}